# Inverse currents in Coulomb-coupled quantum dots

Yanchao Zhang[*], Zhenzhen Xie

*School of Science, Guangxi University of Science and Technology, Liuzhou 545006, People's Republic of China*

The inverse current, i.e., induced current is opposite to applied force, has recently been found in a classical one-dimensional interacting Hamiltonian system [Phys. Rev. Lett. 124, 110607 (2020)]. In this paper, we show that inverse current also exists in quantum system. Based on Coulomb-coupled quantum dots system, we find that inverse current will appear when Coulomb interaction increases. This does not violate the second law of thermodynamics, since entropy reduction caused by inverse current is compensated by entropy increase caused by forward current, which ensure that total entropy increase of the system is always greater than zero.

---

[*] Email: zhangyanchao@gxust.edu.cn

## I. INTRODUCTION

In equilibrium thermodynamics, the response of a system at thermal equilibrium to an applied force is always in the same direction as the applied force in order to move towards a new equilibrium. A counterintuitive behavior called absolute negative mobility (ANM) that system's response is opposite to the applied force is impossible, otherwise it will violate the second law of thermodynamics [1-4]. However, in nonequilibrium systems, ANM can appear and has aroused great interest in recent decades. ANM was originally thought to be merely the result of quantum effects [5,6]. Recently, ANM has also exhibited in a class of classical systems consisting of models of various interacting Brownian particles [1-4, 7-14], and has even been found in different systems both theoretically [15, 16] and experimentally [17].

In coupled transport systems, inverse current, i.e., one induced current (e.g., either particle or energy) against both applied forces, is highly counterintuitive, though there is no fundamental law that forbids inverse current. This means, under appropriate conditions, inverse currents in coupled transport (ICC) can appear in equilibrium system applied with two thermodynamic forces [18]. Indeed, it has been shown in an abstract stochastic model [19]. However, this is an abstract model rather than a purely physical system. Recently, Wang *et al.* determined that ICC can occur in a purely physical system, and their study shown that ICC of energy and particle can occur in a classical one-dimensional interacting Hamiltonian system when its equilibrium state is perturbed by coupled thermodynamic forces [18]. Here, we will show that inverse current also exists in quantum system.

In this paper, we investigate the coupled transport in Coulomb-coupled quantum dots system, which have been extensively studied in thermoelectricity [19-26], thermal rectification [27-33], quantum information [34-37], and thermometry [38,39]. We show that inverse currents, either particle current or energy current, can take place in Coulomb-coupled quantum dots system, and Coulomb interaction is necessary to achieve inverse current. The inverse current does not violate the second law of thermodynamics, because entropy reduction caused by inverse current is compensated by entropy increase caused by forward current.

## II. MODEL AND THEORY

The Coulomb-coupled quantum dots system is illustrated in Fig. 1, where two quantum dots, denoted by $QD_t$ (top) and $QD_b$ (bottom), are coupled to two reservoirs at temperature $T_\nu$ and chemical potential $\mu_\nu$ via tunnel contacts $\gamma_t^\nu$ and $\gamma_b^\nu$ ($\nu = L, R$). The two quantum dots are capacitively coupled, where only exists energy exchange but no particle exchange. The strength of the coupling is characterized by Coulomb energy $U$, which regulates the energy exchange between two quantum dots [19].

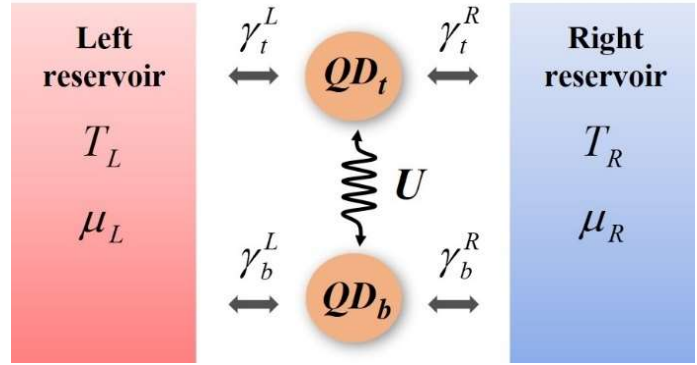

FIG. 1. The schematic diagram of a Coulomb-coupled quantum-dot system, which consisted of two capacitively coupled quantum dots connected to two reservoirs with different temperatures and chemical potentials in the Coulomb-blockade regime.

In the Coulomb blockade regime, each quantum dot with a spinless single energy level $\varepsilon_t$ and $\varepsilon_b$ can be occupied by up to one electron. So the system is given by four states labeled as $|0,0\rangle$, $|1,0\rangle$, $|0,1\rangle$, and $|1,1\rangle$, where 0 or 1 represents that the sites of $\varepsilon_t$ and $\varepsilon_b$ is empty or filled, respectively. In the weak coupling limit, the dynamics of system is well described by a master equation $d\bm{p}/dt = \bm{M}\bm{p}$ for the reduced density matrix $\bm{p} = (p_{00}, p_{10}, p_{01}, p_{11})$ describing the occupation probabilities for four states [19-21]. The $M$ denotes the matrix containing the transition rates and is given by

$$M = \begin{pmatrix} -\sum_{\alpha,v}\Gamma^{v+}_{\alpha 0} & \sum_{v}\Gamma^{v-}_{t 0} & \sum_{v}\Gamma^{v-}_{b 0} & 0 \\ \sum_{v}\Gamma^{v+}_{t 0} & -\sum_{v}\Gamma^{v-}_{t 0}-\sum_{v}\Gamma^{v+}_{b 1} & 0 & \sum_{v}\Gamma^{v-}_{b 1} \\ \sum_{v}\Gamma^{v+}_{b 0} & 0 & -\sum_{v}\Gamma^{v-}_{b 0}-\sum_{v}\Gamma^{v+}_{t 1} & \sum_{v}\Gamma^{v-}_{t 1} \\ 0 & \sum_{v}\Gamma^{v+}_{b 1} & \sum_{v}\Gamma^{v+}_{t 1} & -\sum_{\alpha,v}\Gamma^{v-}_{\alpha 1} \end{pmatrix},\qquad(1)$$

where $\Gamma^{v\pm}_{\alpha n} = \gamma^{v}_{\alpha n} f^{\pm}_{v}(\varepsilon_{\alpha} + U_{\alpha n})$ describe tunneling rates that take an electron into (+) or out (-) of the quantum dot $\alpha$ ($\alpha = t, b$) through reservoir $v$ when the occupation number of other quantum dot is $n$ ($n=0,1$). $\gamma^{v}_{\alpha n}$ is the bare tunneling rate, which describes the coupling strength between quantum dots and reservoirs. $f^{+}_{v}(x) = [1 + e^{(x-\mu_v)/k_B T_v}]^{-1}$ is the Fermi-Dirac function and $f^{-}_{v}(x) = 1 - f^{+}_{v}(x)$. $\varepsilon_{\alpha}$ is the bare energy of single level in the quantum dot $\alpha$. $U_{\alpha n}$ is the charging energy of quantum dot $\alpha$, depending on the occupation number $n$ of other quantum dot. Details on the charging energies of Coulomb-coupled quantum dots system are given in Supplemental Material [40].

From the stationary solution of the master equation, i.e., $Mp = 0$, the steady-state particle (charge) current is given by

$$J_{\rho} \equiv J_{L} = J^{L}_{t} + J^{L}_{b},\qquad(2)$$

where $J^{L}_{t}$ and $J^{L}_{b}$ come from the contributions of $QD_t$ and $QD_b$, respectively, and are given as

$$J^{L}_{t} = q\sum_{n}\left(\Gamma^{L+}_{tn} p_{0n} - \Gamma^{L-}_{tn} p_{1n}\right),\qquad(3)$$

and

$$J^{L}_{b} = q\sum_{n}\left(\Gamma^{L+}_{bn} p_{n0} - \Gamma^{L-}_{bn} p_{n1}\right).\qquad(4)$$

The steady-state energy (heat) current is given by

$$J_u \equiv J_R = J_t^R + J_b^R, \tag{5}$$

with

$$J_t^R = \sum_n \left(\varepsilon_t + U_t^R + \delta_{1n}U\right)\left(\Gamma_{tn}^{R-} p_{1n} - \Gamma_{tn}^{R+} p_{0n}\right) \tag{6}$$

and

$$J_b^R = \sum_n \left(\varepsilon_b + U_b^R + \delta_{1n}U\right)\left(\Gamma_{bn}^{R-} p_{n1} - \Gamma_{bn}^{R+} p_{n0}\right), \tag{7}$$

where $U_t^R = U_{t0} - \mu_R$, $U_b^R = U_{b0} - \mu_R$ [40], and $\delta_{11}=1$ when $n=1$, and $\delta_{10}=0$ when $n=0$.

### III. INVERSE CURRENTS

In the following discussion, we set $\varepsilon = 0$ as the energy reference point. The single energy $\varepsilon_t = \varepsilon + \Delta\varepsilon/2$ and $\varepsilon_b = \varepsilon - \Delta\varepsilon/2$ with $\Delta\varepsilon = \varepsilon_t - \varepsilon_b$ being the energy level difference. We set $T_L = T + \Delta T/2$ and $T_R = T - \Delta T/2$, where $T = (T_L + T_R)/2$ is the average temperature of the two reservoirs and $\Delta T = T_L - T_R$ is the temperature gradient. $q\Delta V = qV_L - qV_R = \mu_L - \mu_R$ is the voltage bias. The other parameters: $\gamma_{\alpha n}^\nu = \gamma$, $q^2/C_\alpha = 20\hbar\gamma$, $k_B T = 7.5\hbar\gamma$, and $\Delta\varepsilon = 3\hbar\gamma$.

In Fig. 2, the particle current and energy current are shown as a function of the Coulomb interaction, and all the applied forces, i.e., temperature gradient and voltage bias, are always non-negative. It can be clearly seen that as Coulomb interaction increases, the inverse currents appear, i.e., either particle current or energy current against both applied forces. This indicates that Coulomb interaction is necessary to achieve inverse currents in Coulomb-coupled quantum dots system. It can be seen from Fig. 2(a) and 2(b) that the inverse particle current gradually decreases and eventually disappears with the increase of voltage bias, while it gradually increases with the increase of temperature gradient. As shown in Fig. 2(c) and 2(d), the inverse energy current increases with the increase of voltage bias, and gradually decreases and eventually disappears with the increase of temperature gradient.

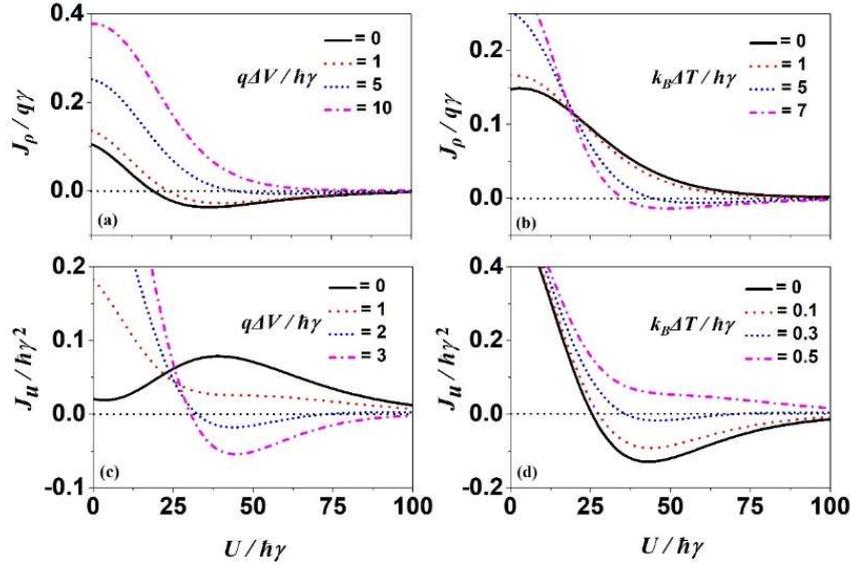

FIG. 2. The particle current $J_\rho$ (a) and (b), and energy current $J_u$ (c) and (d) as a function of the Coulomb interaction for (a) $k_B \Delta T = 5\hbar\gamma$, (b) $q\Delta V = 5\hbar\gamma$, (c) $k_B \Delta T = 0.2\hbar\gamma$ and (d) $q\Delta V = 3\hbar\gamma$.

A map of the particle current as functions of temperature gradient and voltage bias is shown in Fig. 3. It is seen that the inverse particle current occurs in the region bounded by the solid and dotted lines. In the area of $\Delta T > 0$ and $\Delta V > 0$, the particle current $J_\rho < 0$, i.e., the particle current flows from low temperature and low chemical potential to high temperature and high chemical potential, whereas energy current $J_u > 0$. In another area of $\Delta T < 0$ and $\Delta V < 0$, the currents $J_\rho > 0$ and $J_u < 0$.

Fig. 4 shows a map of the energy current as functions of temperature gradient and voltage bias and shows the region where inverse energy current occurs. Within this region the energy current against both applied forces. Fig. 4 also clearly shows that the inverse energy current occurs only when temperature difference is very small. Therefore, it can be seen from Fig. 3 and Fig. 4 that the two inverse currents areas do not overlap. This is to be expected, otherwise it would violate the second law of thermodynamics.

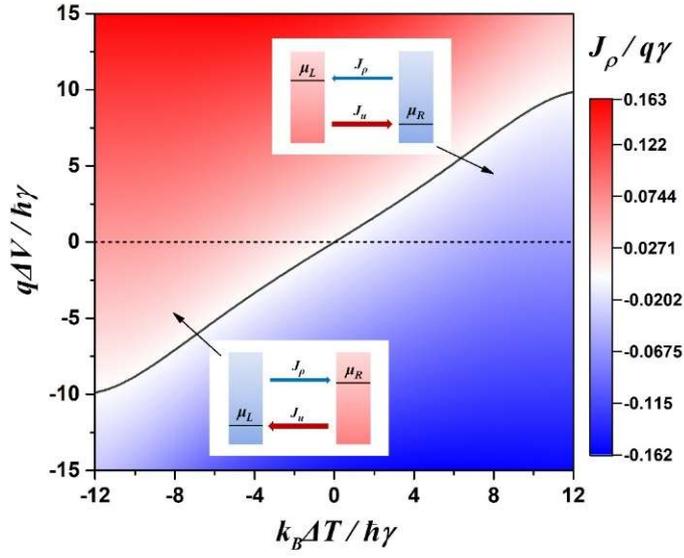

FIG. 3. The particle current $J_\rho$ as functions of the temperature gradient $\Delta T$ and the voltage bias $\Delta V$ for $U/\hbar\gamma = 40$. The inverse particle current occurs in the region bounded by the solid and dotted lines, as shown in inset.

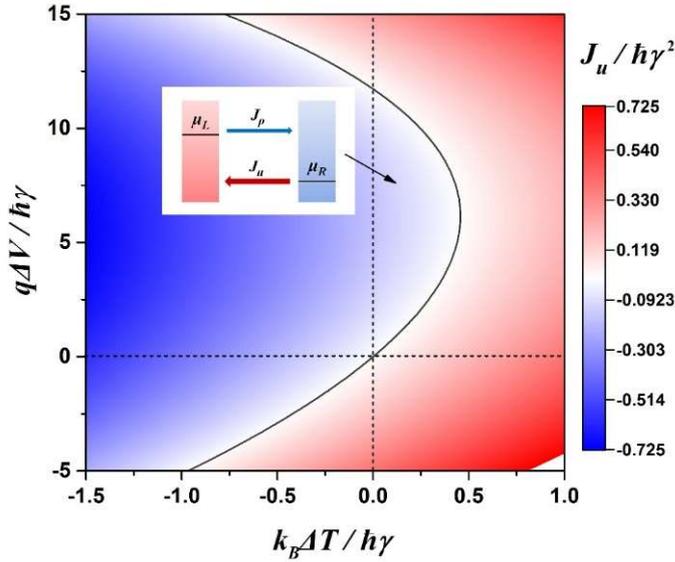

FIG. 4. The energy current $J_u$ as functions of the temperature gradient $\Delta T$ and the voltage bias $\Delta V$ for $U/\hbar\gamma = 40$. The inverse energy current occurs in the region bounded by the solid and dotted lines, as shown in inset.

## IV. ENTROPY CURRENT

In order to understand the inverse currents do not violate the second law of thermodynamics, we set $\Delta T > 0$ and $\Delta V > 0$, so that a negative current means inverse current. Thus, the entropy current of the system, i.e., $J_S = -J_L/T_L + J_R/T_R$ can be rewritten as

$$J_S = \left(-\frac{J_R}{T_L} + \frac{J_R}{T_R}\right) + \left(-\frac{J_L - J_R}{T_L}\right), \tag{8}$$

According to the inverse current shown in Fig. 4, the energy current $J_u$ acts as an inverse current against both applied forces, so the first term in Eq. (8) represents the entropy current due to the inverse energy current, which is defined as $J_S^r = -J_R/T_L + J_R/T_R$. Based on the first law of thermodynamics, i.e., $-J_L + J_R = J_\rho \Delta V$, this indicates that the second term in Eq. (8) is caused by the forward particle current $J_\rho$ and is expressed as $J_S^f = -(J_L - J_R)/T_L$.

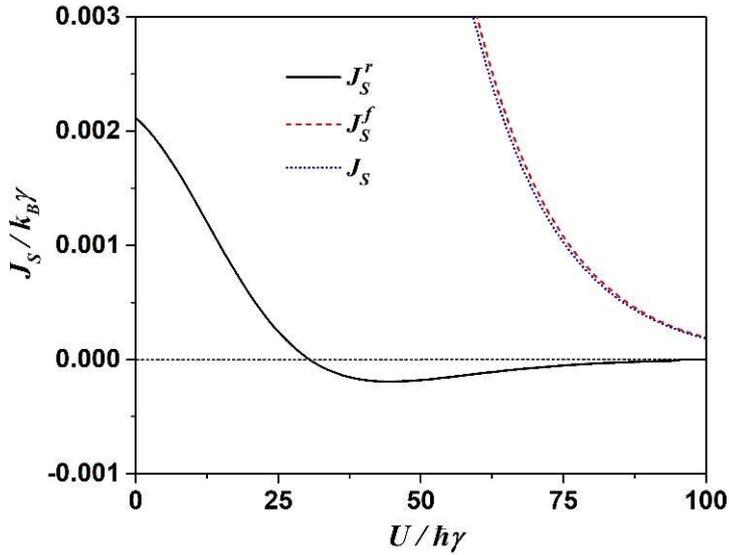

FIG. 5. The entropy current $J_S^r$, $J_S^f$ and $J_S$ as a function of the Coulomb interaction for

$$k_B \Delta T = 0.2\hbar\gamma \text{ and } q\Delta V = 3\hbar\gamma.$$

In Fig. 5 the entropy current $J_S^r$, $J_S^f$ and $J_S$ are shown as a function of the Coulomb interaction for $k_B \Delta T = 0.2\hbar\gamma$ and $q\Delta V = 3\hbar\gamma$. It can be clearly found that as Coulomb interaction increases, the entropy current $J_S^r$ will appear negative entropy, and the inverse current will also appear at this time, as shown by the magenta curve in Fig. 2(c). However, the entropy current $J_S^f$ caused by forward current is much larger than the entropy current $J_S^r$ caused by inverse current, and the negative entropy caused by inverse current is compensated to ensure that total entropy current $J_S$ of the system is always greater than zero.

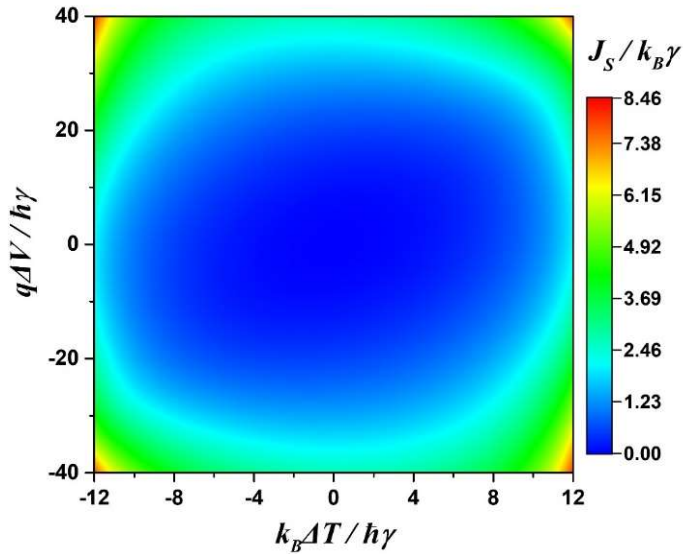

FIG. 6. The entropy current $J_S$ as functions of the temperature gradient $\Delta T$ and the voltage bias $\Delta V$ for $U/\hbar\gamma = 40$.

In order to have an overall grasp of entropy current, a map of the entropy current as functions of temperature gradient and voltage bias is shown in Fig. 6. As expected, entropy current $J_S$ is always greater than zero in the whole parameters range, satisfying the second law of thermodynamics.

9 / 15

## V. CONCLUSIONS

We have shown that inverse currents, either particle current or energy current, can take place in Coulomb-coupled quantum dots system. The results show that Coulomb interaction is necessary to achieve inverse current. We have also demonstrated that inverse current does not violate the second law of thermodynamics, because entropy reduction caused by inverse current is compensated by entropy increase caused by forward current. These results may stimulate greater interest in micro-nano-scale coupled transport, and develop new effects in the fields of rectification and thermoelectricity.

## ACKNOWLEDGEMENTS

This paper is supported by the National Natural Science Foundation of China (No. 11947010), the Science and Technology Base and Talent Project of Guangxi (No. AD19110104)

# Supplemental material: Inverse currents in Coulomb-coupled quantum dots

Yanchao Zhang, Zhenzhen Xie

*School of Science, Guangxi University of Science and Technology, Liuzhou 545006, People's Republic of China*

**Charging energies of a Coulomb-coupled quantum dots system**

The electrostatic model of two capacitively coupled quantum dots $QD_t$ and $QD_b$, connected to two reservoirs $v$ ($v = L, R$) is sketched in Fig. 1, The capacitances between the quantum dot $\alpha$ ($\alpha = t, b$) and the reservoirs $v$ and between $QD_t$ and $QD_b$ are defined as $C_\alpha^v$ and $C$, respectively. $\phi_\alpha$ is the electrostatic potential of quantum dot $\alpha$ and $V_v$ is the bias voltage of the reservoirs $v$.

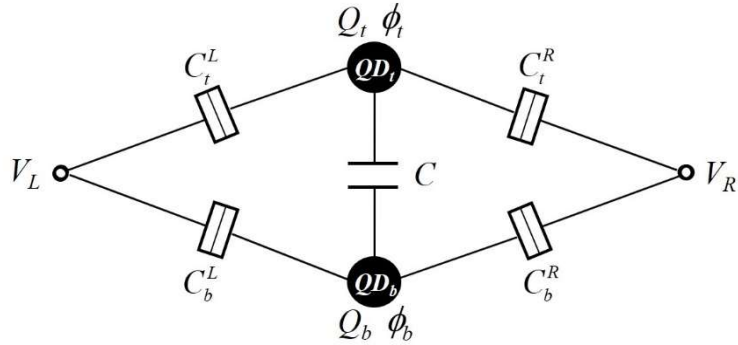

FIG. 1. The electrostatic model of a Coulomb-coupled quantum dots connected to two reservoirs.

The electrostatic equations for the charges $Q_t$ and $Q_b$ are given by [1]

$$Q_t = C_t^L(\phi_t - V_L) + C_t^R(\phi_t - V_R) + C(\phi_t - \phi_b), \tag{1}$$

$$Q_b = C_b^L(\phi_b - V_L) + C_b^R(\phi_b - V_R) + C(\phi_b - \phi_t). \tag{2}$$

A matrix form can be expressed as

$$\begin{pmatrix} Q_t + C_t^L V_L + C_t^R V_R \\ Q_b + C_b^L V_L + C_b^R V_R \end{pmatrix} = \begin{pmatrix} C_{\Sigma t} & -C \\ -C & C_{\Sigma b} \end{pmatrix} \begin{pmatrix} \phi_t \\ \phi_b \end{pmatrix}, \tag{3}$$

where $C_{\Sigma t} = C_t^L + C_t^R + C$ and $C_{\Sigma b} = C_b^L + C_b^R + C$ define the total capacitances of $QD_t$ and $QD_b$. The electrostatic potentials $\phi_t$ and $\phi_b$ are then expressed as

$$\begin{pmatrix} \phi_t \\ \phi_b \end{pmatrix} = \frac{1}{C_{\Sigma t} C_{\Sigma b} - C^2} \begin{pmatrix} C_{\Sigma t} & C \\ C & C_{\Sigma b} \end{pmatrix} \begin{pmatrix} Q_t + C_t^L V_L + C_t^R V_R \\ Q_b + C_b^L V_L + C_b^R V_R \end{pmatrix}. \tag{4}$$

Thus, the electrostatic energy for state of system is given by

$$U(n_t, n_b) = \frac{1}{2} \begin{pmatrix} Q_t + C_t^L V_L + C_t^R V_R \\ Q_b + C_b^L V_L + C_b^R V_R \end{pmatrix}^T \begin{pmatrix} \phi_t \\ \phi_b \end{pmatrix}, \tag{5}$$

where $Q_\alpha = n_\alpha q$ and $n_\alpha$ is electron number of quantum dot $\alpha$. In the Coulomb blockade regime, each of quantum dots can be occupied by up to one electron ($n_\alpha = 0, 1$). Thus, the electrostatic energies for the four states read $U(0,0)$, $U(1,0)$, $U(0,1)$, and $U(1,1)$. The charging energy of quantum dot $\alpha$ is defined by $U_{\alpha n}$, depending on the occupation number $n$ of the other quantum dot. When an electron tunnels into a quantum dot while the other dot is empty, the charging energies of $QD_t$ and $QD_b$ are, respectively, given by

$$U_{t0} = U(1,0) - U(0,0), \tag{6}$$

$$U_{b0} = U(0,1) - U(0,0). \tag{7}$$

However, when the other quantum dot is occupied, the charging energies are, respectively, given by

$$U_{t1} = U(1,1) - U(0,1) = U_{t0} + U, \tag{8}$$

$$U_{b1} = U(1,1) - U(1,0) = U_{b0} + U, \tag{9}$$

where

$$U = \frac{q^2 C}{C_{\Sigma t} C_{\Sigma b} - C^2} \tag{10}$$

depends on the capacitance of the system and determines the exchanged energy between the two quantum dots when an electron tunnels into the empty quantum dot but leaves it only after a second electron has occupied the other quantum dot [2]. Here, we set

$C_t^L = C_t^R = C_b^L = C_b^R \equiv C_\alpha$, one has

$$U = \frac{q^2 C}{4C_\alpha (C_\alpha + C)}, \tag{11}$$

and

$$U_t^L = U_b^L = \frac{q^2 (2C_\alpha + C) - 4qC_\alpha (C_\alpha + C)(V_L - V_R)}{8C_\alpha (C_\alpha + C)}, \tag{12}$$

$$U_t^R = U_b^R = \frac{q^2 (2C_\alpha + C) + 4qC_\alpha (C_\alpha + C)(V_L - V_R)}{8C_\alpha (C_\alpha + C)}, \tag{13}$$

where $U_t^L = U_{t0} - \mu_L$, $U_t^R = U_{t0} - \mu_R$, $U_b^L = U_{b0} - \mu_L$, and $U_b^R = U_{b0} - \mu_R$.